\documentclass[preprint,review,3p,12pt]{elsarticle}

\usepackage{amssymb}
\usepackage{amsmath}
\setcitestyle{square}
\setcitestyle{citesep={,}}
\usepackage[colorlinks=true,linkcolor=blue,allcolors=blue,citecolor=blue]{hyperref}
\setcitestyle{number}
\journal{Computational Materials Science}

\begin{document}

\begin{frontmatter}



\title{Interface Response Functions for multicomponent alloy solidification- An application to additive manufacturing}


\author[inst1]{V S Hariharan\texorpdfstring{\corref{cor1}}}
\ead{rajhharan97@gmail.com}
\cortext[cor1]{Corresponding Author}
\affiliation[inst1]{organization={Department of Metallurgical and Materials Engineering, Indian Institute of Technology Madras},%Department and Organization
            %%addressline={Address One}, 
            city={Chennai},
            postcode={600036}, 
            state={Tamil Nadu},
            country={India}}

\author[inst1,inst2]{B S Murty}
\author[inst1]{Gandham Phanikumar}
%%\ead{gphani@iitm.ac.in}
\affiliation[inst2]{organization={Indian Institute of Technology Hyderabad},%Department and Organization
           %% addressline={Address Two}, 
            city={Kandi},
            postcode={502284}, 
            state={Telangana},
            country={India}}

\begin{abstract}
The near-rapid solidification conditions during additive manufacturing can lead to selection of non-equilibrium phases. Sharp interface models via interface response functions have been used earlier to explain the microstructure selection under such solidification conditions. However, most of the sharp interface models assume linear superposition of contributions of alloying elements without considering the non-linearity associated with the phase diagram. In this report, both planar and dendritic Calphad coupled sharp interface models have been implemented and used to explain the growth-controlled phase selection observed at high solidification velocities relevant to additive manufacturing. The implemented model predicted the growth-controlled phase selection in multicomponent alloys, which the other models with linear phase diagram could not. These models are calculated for steels and a Nickel-based superalloy and the results are compared with experimental observations.
\end{abstract}

\begin{graphicalabstract}
\includegraphics[width=14cm]{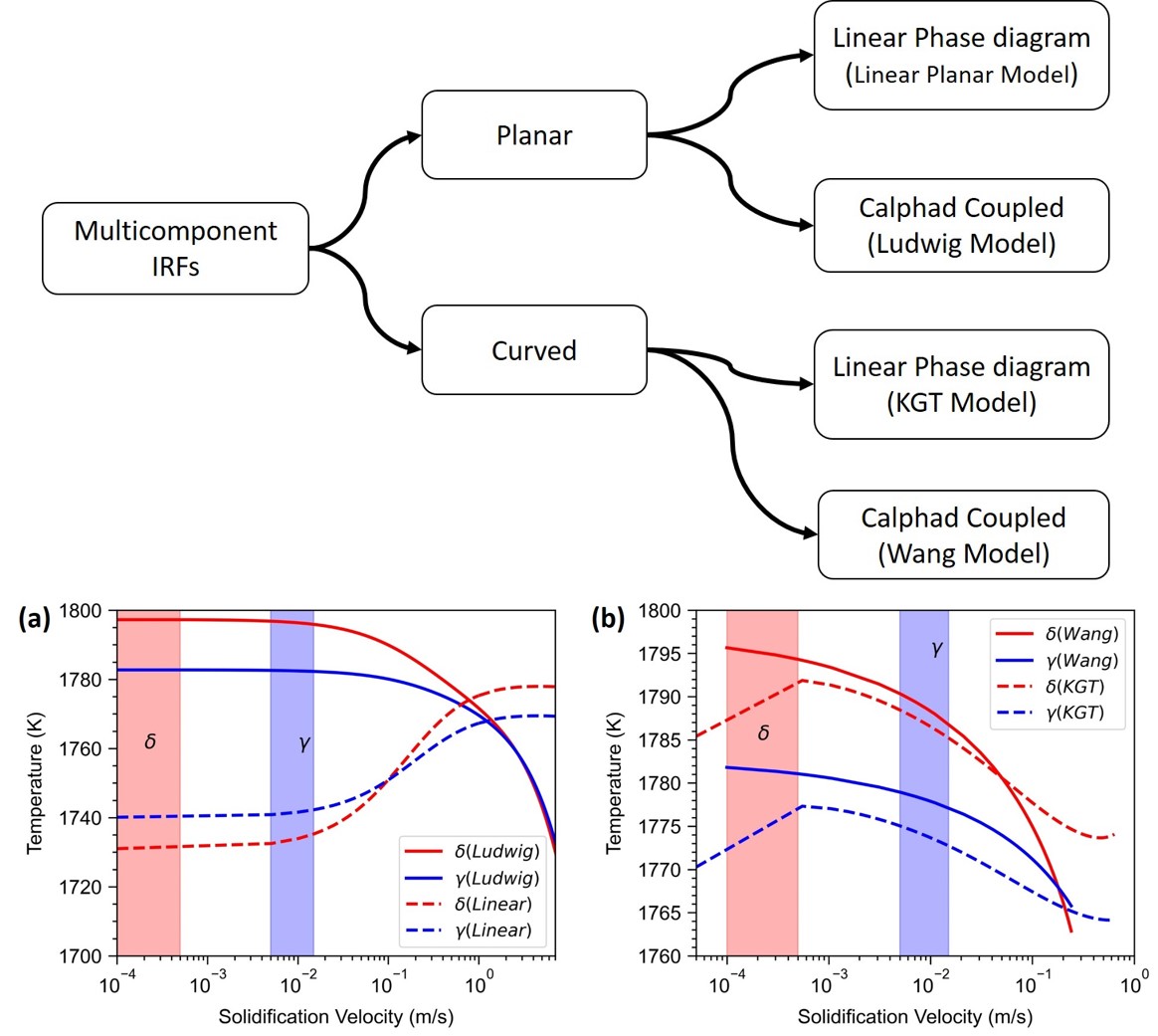}
\centering
\end{graphicalabstract}

\begin{highlights}
\item Calphad coupled dendritic growth model predicted the growth-controlled phase selection occurring during additive manufacturing of steels, that is not predicted by other growth models
\item Incorporating the effect of solute drag improved the predictions
\item Columnar to equiaxed transition (CET) calculation is coupled with KGT model and benchmarked with experiments for Haynes 282 Ni-based superalloy
\end{highlights}

\begin{keyword}
Additive Manufacturing\sep Calphad \sep Microstructure \sep Steels \sep Superalloys
%%\PACS 0000 \sep 1111
%%\MSC 0000 \sep 1111
\end{keyword}

\end{frontmatter}


\section{Introduction}
Additive Manufacturing (AM) involves building three-dimensional parts layer-by-layer based on a digital model. AM has proven its ability to produce complex, near-net shape components making it appealing for biomedical and aerospace applications, especially to manufacture high-value low-volume components \cite{debroy2018additive}. The near-rapid solidification condition and the thermal cycles experienced during additive manufacturing can lead to non-equilibrium microstructures \cite{babu2018additive}. Understanding the microstructure evolution during AM can help in alloy design for AM as well as site-specific control of microstructure within a component \cite{dehoff2015site}.\par 

The solidification velocities in AM processes like electron powder bed fusion can reach up to 10 cm/s \cite{raghavan2017localized}.  Equilibrium thermodynamic calculations alone are insufficient to understand the microstructure evolution during such rapid solidification conditions. One must account for the deviation from equilibrium at the solid-liquid interface.  Analytical solidification models, also known as sharp interface models have been developed to explain the solute segregation behaviour and phase selection during processes such as welding and directional solidification \cite{kurz2001solidification}. The sharp interface models assume that the solidification occurs by an “interface reaction” at the solid-liquid interface and there is long-range transport of atoms and heat across the interface. Interface response functions appropriately describe the interface reaction. The interface response functions (IRFs) can be used to predict the solid composition and the interface velocity for a given composition and temperature of the interface or vice-versa.\par

The concept of the IRF is developed using the principles of irreversible thermodynamics \cite{BakerCahn}. IRFs have been used in the solidification and solid-state transformations. We restrict ourselves to the context of solidification for this study. As given by Hillert, at the interface, Gibbs energy dissipation occurs due to crystallization and diffusion of solute species across the interface (trans-interface diffusion) \cite{hillert1999solute}. The relationship between the driving forces and their corresponding fluxes at the interface dictates the IRFs . One such relationship provided by Aziz and Kaplan in their continuous growth model gives the relationship between partition coefficient and solidification velocity \cite{aziz1988continuous}. They also show how the phase diagram varies with solidification velocity (kinetic phase diagram) \cite{aziz1988continuous}.

Aziz and Kaplan model considers only the planar interface. In order to arrive at IRF for curved interfaces, one must know the conditions at which the planar interface becomes unstable. Mullins and Sekerka derived the condition for planar stability in the light of perturbation analysis for low Peclet numbers \cite{mullins1964stability}. Based on this work, Langer and M\"{u}ller-Krumbharr proposed the Marginal Stability criterion which assumes that the marginally stable wavelength gets selected as the dendritic tip radius \cite{langer1978theory}. Trivedi and Kurz extended Mullins and Sekerka's analysis to rapid solidification conditions \cite{trivedi1986morphological}. Building on the above models, Kurz et al. have proposed the IRF for rapid solidification, where one can obtain the interface composition and temperature as a function of solidification velocity and thermal gradient \cite{kurz1986theory}. This model is popularly referred to as KGT model, and is widely used in the solidification community.

The KGT model has been used to explain the growth-controlled phase selection in Fe-Cr-Ni alloys \cite{fukumoto1997delta}. Fukumoto and Kurz observed that equilibrium delta ferrite ($\delta$ BCC) is the primary solidification phase for low solidification velocities and austenite ($\gamma$ FCC) is the primary solidification phase for high solidification velocities \cite{fukumoto1997delta}. This phase selection was explained using interface temperature versus solidification velocity plot obtained using KGT model. The phase with higher interface temperature gets selected and since the interface temperature of $\gamma$ is higher than $\delta$ at high solidification velocities, $\gamma$ is preferred over $\delta$ \cite{fukumoto1997delta}. Several peritectic systems of commercial interest such as steels and Nd-Fe-B alloys show this kind of growth-controlled phase selection \cite{kerr1996solidification}. Although KGT model was originally developed for binary alloys, it has been used for multicomponent alloys by assuming a simple linear superposition of individual solute contributions \cite{babu2002time}. This approximation can lead to errors in the predictions.

Ludwig solved this issue by extending Aziz and Kaplan’s continuous growth model to multicomponent alloys by considering the thermodynamic effects instead of linear superposition \cite{ludwig1998interface}.  Recently, Du et al. have reformulated Aziz and Kaplan’s model in terms of Gibbs energy instead of the original chemical potentials and coupled with Calphad (Calculation of Phase Diagram) databases \cite{du2022kinetic}.  The model was applied in the context of additive manufacturing for a multicomponent aluminium alloy. The reformulated model by Du et al. and Ludwig’s model are similar and yield same results. However, Ludwig’s model is not widely applied due to its increased computational complexity, except for a few studies \cite{wang2013planar}. These models are valid only for planar interface. Wang et al. proposed a Calphad coupled model for dendritic growth that considers non-Fickian diffusion in liquid, which will be relevant for rapid solidification conditions experienced during additive manufacturing \cite{wang2013dendritic}. However, these models are applied only for undercooling \cite{wang2013dendritic} and casting conditions \cite{wu2023generalized}. 

In this article, we couple Ludwig’s model and Wang's model for multicomponent alloy solidification with Calphad databases and explain the growth-controlled phase selection observed during additive manufacturing. The implemented model is compared with the existing analytical solidification models that are based on linear superposition, in terms of its predictive capability and the relevant physics associated. As case studies, we take commercially relevant two steel compositions (Fe-C-Mn-Al steel, H13 tool steel) processed under welding \cite{babu2002time} and additive manufacturing \cite{konig2023solidification} conditions, for which in situ synchrotron studies are available in the literature to determine the primary solidification phase. In addition to phase selection, the prediction of columnar to equiaxed transition also becomes important in the case of additive manufacturing of Nickel-based superalloys \cite{kurz2001columnar}. In relation to the previous models, we also calculate columnar to equiaxed transition map for Haynes 282 and the impact of model assumptions are discussed.

\section{Models}
\begin{figure}[h]
\includegraphics[width=14cm]{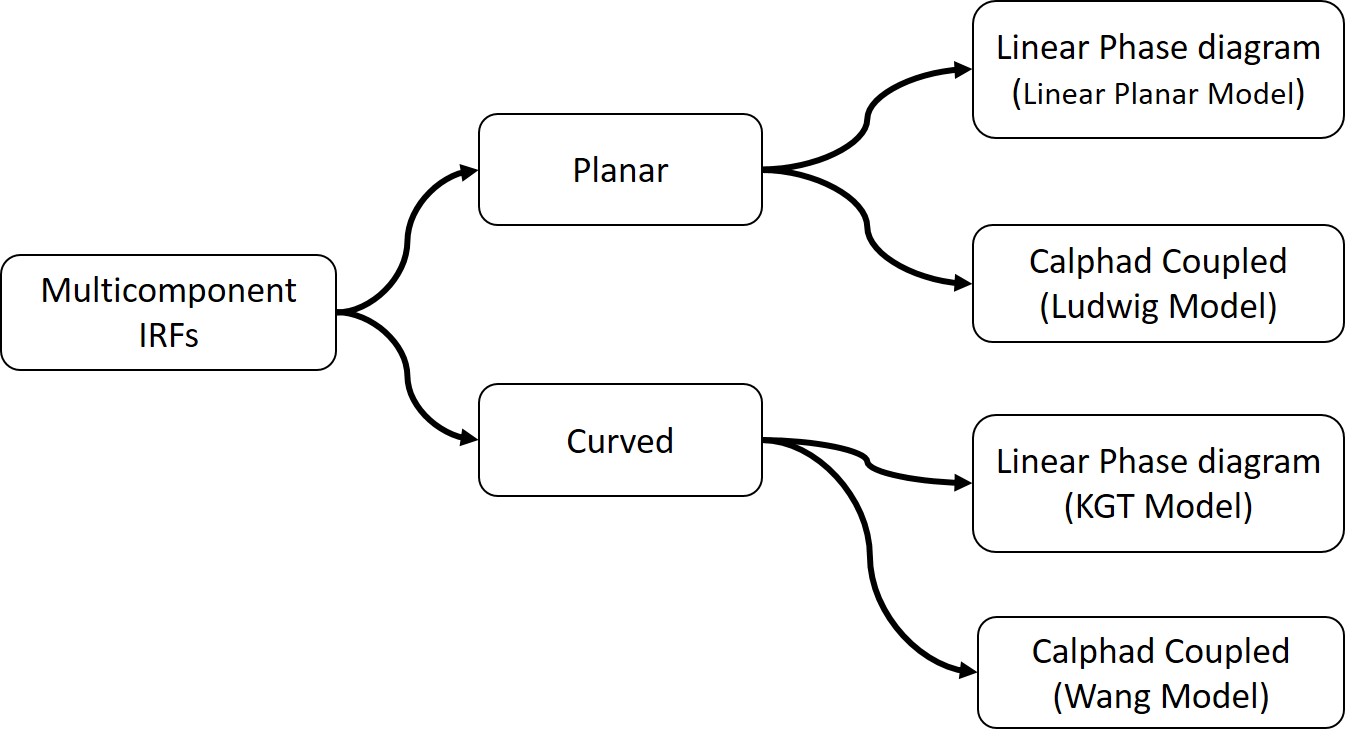}
\centering
\caption{Types of Interface Response Functions (IRFs) used in this work}
\label{fig:IRF_types}
\end{figure}

The IRFs can vary depending on the model and its assumptions used. In this report, we considered both planar and curved interface models that use linear phase diagram and the ones that are calphad coupled. The models used in this work are shown in Fig.\ref{fig:IRF_types}

These models assume one-dimensional and steady state growth conditions on the heat and mass transfer scale that occurs during solidification. Regarding AM, due to the metal-laser interaction, there is a spatial and temporal variation of temperature within the melt pool. However, the models are applied on the scale of solute diffusion length (Solute Diffusivity/solidification velocity $\approx$ 10\textsuperscript{-9} m\textsuperscript{2}s\textsuperscript{-1}/ 10\textsuperscript{-2} m s\textsuperscript{-1} = 10\textsuperscript{-7} m), which is short in the scale of the melt pool. Even though further research is needed to test this assumption, it is plausible to assume that the analytical models can be used to predict microstructure selection during AM.

\subsection{Planar growth with linear phase diagram}
The interface response (temperature) for a planar solid-liquid interface by assuming linear superposition of the individual solute contributions is given by Eqn.\ref{Eqn:planar-linear}.

\begin{align}
    \label{Eqn:planar-linear}
    T_P&=T_m+\sum_iC^i_o\left(\frac{m^i_v}{k^i_v}\right)-\frac{v}{\mu_k}\\
    \label{Eqn:kinetic-k}
    k^i_v&=\frac{k^i_o+\frac{v}{v_D}}{1+\frac{v}{v_D}}\\
    \label{Eqn:kinetic-m}
    m^i_v&=m^i_o\times\frac{1-k^i_v(1-\ln(k^i_v/k^i_o))}{1-k^i_o}   
\end{align}

In Eqn.\ref{Eqn:planar-linear},  $T_P$ is the temperature of the planar interface, $T_m$ is the melting point of the pure metal, the summation over the alloying element ‘$i$’ allows for extension to multicomponent alloys, $C^i_o$ represents the alloy composition, $m^i_o$ represents the equilibrium liquidus slope, $k^i_o$ represents equilibrium partition coefficient, the letter ‘$v$’ in the subscript represents that the terms are velocity-dependent. The velocity-dependent partition coefficient and liquidus slope are given in Eqn.\ref{Eqn:kinetic-k} \cite{aziz1988continuous} and Eqn.\ref{Eqn:kinetic-m} respectively. ‘$v$’ represents the solidification velocity, ‘$\mu_k$’ represents the kinetic coefficient, '$v_D$' is the diffusive speed of elements in liquid, which is assumed to be the same as that of the solvent. The velocity dependent partition coefficient is from Aziz and Kaplan model \cite{aziz1988continuous}. The interface temperature can be obtained as shown in Eqn.\ref{Eqn:planar-linear} only by assuming that Henry's law holds and the solution is dilute \cite{BCTmodel,ludwig1998interface}.  Diffusion in solid is assumed negligible.

\subsection{Dendritic growth with linear phase diagram}
The interface response of curved interface based on KGT model is given in Eqn. 4 \cite{kurz1986theory}. Again, a linear superposition of individual solutes is assumed when extended to multicomponent alloys \cite{fukumoto1997delta,babu2002time}.

\begin{align}
\label{Eqn:KGT}
    &T_D=T_L+\sum_i(C^i_Lm^i_v-C^i_om^i_o)-\frac{2\Gamma}{R}-\frac{v}{\mu_k}-\frac{GD_i}{v}\\
    &C^i_L=\frac{C^i_o}{1-\left((1-k^i_v) Iv(Pe^i)\right)}\\
    \label{Eqn:KGT-radius}
    &4\pi^2\Gamma\left(\frac{1}{R^2}\right)+(2\sum_i\left[m^i_vPe^i(1-k^i_v)C^i_L\xi^i_C\right])(\frac{1}{R})+G=0\\
    &\xi^i_C=1-\frac{2k^i_v}{2k^i_v-1+\sqrt{1+{\left( \frac{2\pi}{Pe^i} \right)}^2}}
\end{align}

In Eqn.\ref{Eqn:KGT}, $T_D$ is the temperature of the curved interface, $T_L$ is the equilibrium liquidus temperature of the alloy, the second term in right-hand side of the equation represents solutal contribution, the third and the fourth term represents the contribution due to curvature of dendritic tip and attachment kinetic effects respectively. The last term accounts for the cellular growth at low solidification velocities. $C^i_L$ is the liquid composition at the interface corrected for the solidification velocity and curvature at the tip. $\rm{Iv}(Pe)= Pe\times\exp(Pe)\times \rm{E1}(Pe)$  is the Ivantsov solution, to describe the solute diffusion field ahead of the dendrite , where E1 is the exponential integral and is calculated using \textit{'scipy'} library \cite{2020SciPy-NMeth}. Ivantsov solution is a function of the dimensionless solutal Peclet number ($Pe^i=VR/2D^i$), which is the ratio of advective solutal transport to diffusive solutal transport, at the scale of dendrite tip radius ($R$). The expressions for velocity-dependent partition coefficient and liquidus slope remain the same for both planar and curved interfaces. Equation \ref{Eqn:KGT-radius} is solved iteratively for a given thermal gradient (G) and a range of solidification velocities to obtain $R$ and then $T_D$ is subsequently calculated. The bisection method was used to solve Eqn.\ref{Eqn:KGT-radius} \cite{haines2018sensitivity}. 

\subsection{Calphad coupled planar growth}
Solidification occurs via the motion of the solid-liquid interface, for the interface to move, there must be a driving force. This driving force is due to the deviation from equilibrium at the interface. This deviation leads to a redistribution of solute across the interface and migration of the interface. Building on the above assumption and Aziz – Kaplan model, Ludwig \cite{ludwig1998interface} derived the first response function given in Eqn.\ref{Eqn:Ludwig-IRF1}.

\begin{align}
\label{Eqn:Ludwig-IRF1}
    &(C^j_L-C^j_S)\left(\frac{v}{v_D}\right)=\sum_{i=1}^n(C^i_LC^j_S-\kappa^{i,j}C^i_SC^j_L)\\
    &\kappa^{i,j}=\exp{\left(-\frac{\Delta\tilde{\mu^j}-\Delta\tilde{\mu^i}}{RT}\right)}\\
    &\Delta\tilde{\mu^i}=\Delta\mu^i-RT\ln(\frac{C^i_S}{C^i_L})\\
    &\tilde{\mu^i}=\mu^i-RT\ln(C^i)
\end{align}

In Eqn.\ref{Eqn:Ludwig-IRF1}, $C^i_L$ and $C^i_S$ are the composition (mole fraction of ‘$i$’th element) of liquid and solid at the interface. $\kappa^{i,j}$ are the partitioning-parameters between the elements $i,j$, such that $i\neq j$.  To account for the local entropic effects, the difference in the redistribution potentials ($\Delta\tilde{\mu ^i}=\tilde{\mu^i_S}-\tilde{\mu^i_L}$) are considered rather than the chemical potentials. The redistribution potential of component ‘$i$’ is defined as the difference between the chemical potential and ideal mixing entropy contribution.$n$-1 equations are to be solved for a ‘$n$’ component system and the composition of the major element (solvent) can be obtained from the constraint $\sum^n_{j=1}C^j_S=1$. The second response function is based on chemical rate theory. It describes the relationship between the driving force for interface motion and the velocity of the interface as given in Eqn.\ref{Eqn:Ludwig_IRF2}.

\begin{equation}
\label{Eqn:Ludwig_IRF2}
    v=v_o[1-\exp(\frac{\Delta G}{RT})]
\end{equation}

In Eqn.\ref{Eqn:Ludwig_IRF2}, $v_o$ is the maximum velocity that occurs when there is infinite driving force and is generally in the order of speed of sound in liquid.$\Delta G$ is the driving force for solidification. The expression for $\Delta G$ can vary depending on how the solute drag effect is treated. $\Delta G_{DF}$ is the Gibbs energy difference between solid and liquid that causes interface motion.  The driving force is assumed to have two components, one due to the crystallization of atoms and the solute redistribution across the interface({$\Delta G_{DF}=\Delta G_C+\Delta G_D$}). The crystallization and solute distribution energies are given below.
\begin{align}
    \Delta G_C=\sum_{i=1}^{i=n} C^i_L\Delta\mu^i\\
    \label{Eqn:Solute_Drag_Energy}
    \Delta G_D=\sum_{i=1}^{i=n} (C^i_S-C^i_L)\Delta\mu^i
\end{align}

If we assert that the driving force must account for all the solidifying atoms, that is, neglecting solute drag, then $\Delta G_{nsd}=\Delta G_{DF}$ and can be written as in Eqn.\ref{Eqn:Nodrag}.
\begin{equation}
    \label{Eqn:Nodrag}
    \Delta G_{nsd}=\sum^n_{i=1}C^i_S.\Delta\mu^i
\end{equation}

On the other hand, if we assume that a part of the driving force is consumed due to the solute diffusion at the interface, that is, including solute drag, then $\Delta G_{sd}=\Delta G_{C}$ and can be written as in Eqn.\ref{Eqn:Solutedrag}.
\begin{equation}
    \label{Eqn:Solutedrag}
    \Delta G_{sd}=\sum^n_{i=1}C^i_L.\Delta\mu^i
\end{equation}

During calculation, if solute drag is included then $\Delta G=\Delta G_{sd}$ and if solute drag is neglected $\Delta G=\Delta G_{nsd}$. The implications of accounting for the effect of solute drag during multicomponent alloy solidification are investigated in the subsequent sections.

\begin{figure}[h]
\includegraphics[width=14cm]{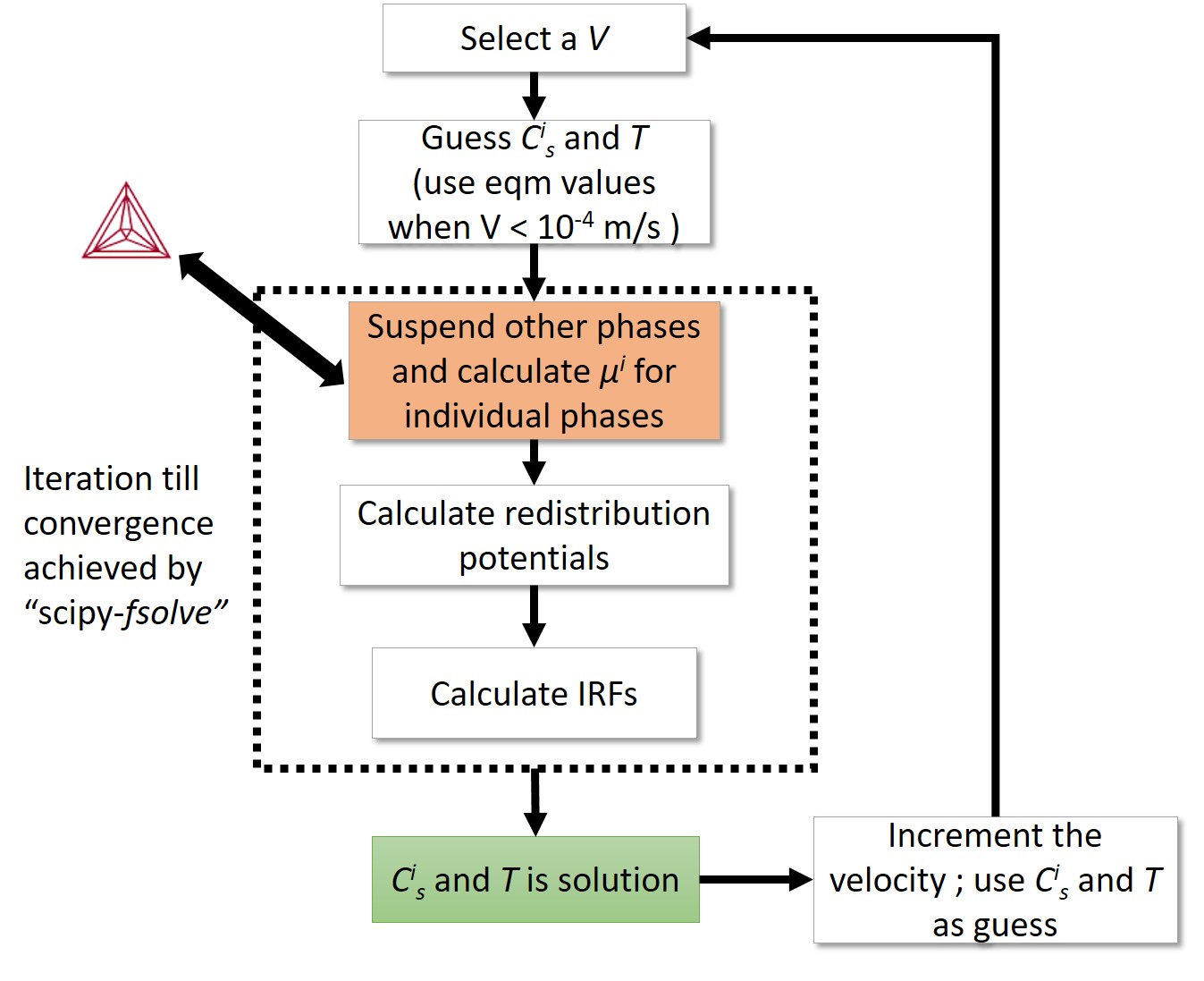}
\centering
\caption{A flowchart to calculate interface temperature and composition using Ludwig model. The orange colour indicates that it is Calphad coupled.}
\label{fig:Ludwig_schematic}
\end{figure}

A flowchart representing the implementation of Ludwig’s model is shown in Fig.\ref{fig:Ludwig_schematic}. Since chemical potentials are needed to solve the system of equations (Eqn. \ref{Eqn:Ludwig-IRF1} -\ref{Eqn:Solutedrag}), TC-python\textsuperscript{\textregistered} API of Thermo-Calc\textsuperscript{\textregistered} software \cite{Thermocalc} is used along with optimization routines in scipy library \cite{2020SciPy-NMeth}. The calculation starts with a very low velocity, so the solid's composition and the interface's temperature are close to the equilibrium values obtained from Calphad databases. These values are used as initial guess values for the '\textit{fsolve}' function in '\textit{scipy}'. While calculating chemical potentials, all the phases are rejected except the phase for which the chemical potential is calculated. Once the solution is converged, these values are stored in an array and used as guess values for the next calculation with higher velocity. 

\subsection{Calphad coupled dendritic growth}

Wang et al proposed a calphad coupled IRF for multicomponent alloy solidification for dendritic growth \cite{wang2013dendritic}. This model (hereafter referred to as Wang model) accounts for local non-equilibrium diffusion in liquid. This results in improved predictions at high solidification velocities. Unlike Ludwig model which requires fixing either solid or liquid composition at the interface, Wang model allows the calculation of temperature, solid and liquid compositions at the interface. The first response functions to calculate the interface compositions are given below.

\begin{align}
\label{eqn:WangIRF1a}
    &\frac{v}{V_m}(C^i_L-C^i_S)=M^i_D\left(\Delta\mu^i\psi^i-\frac{\sum_{j=1}^{n}M^j_D\Delta\mu^j\psi^j}{\sum_{k=1}^nM^k_D}\right)\\
    &M^i_D=\frac{v^i_I}{V_m}\left(\frac{\partial \mu_i^L}{\partial C^L_i}\right)^{-1}
\end{align}

where $V_m$ is the molar volume  and is assumed to be same for all elements. $\psi^i(=1-(v/v^i_D)^2)$ is the non-equilibrium diffusion factor, $M^i_D$ is the mobility for trans-interface diffusion, $v^i_I$ is the diffusive speed at the interface. The above set of equations are same for the planar model proposed by Wang et al \cite{wang2013planar}. In case of dendrite growth, an additional set of equations that accounts for the curvature must be considered. Thus, Ivanstov solutions are used for the same.

\begin{equation}
\label{eqn:WangIRF1b}
    C^i_L= 
\begin{cases}
    \frac{C^i_o}{1-\left((1-k^i_v) Iv(Pe^i)\right)}& \text{if } v < v^i_D\\
    C^i_o              & v \geq v^i_D
\end{cases}
\end{equation}

It can be seen from the above equations that at solidification velocities higher than diffusive speed in bulk liquid, there is no solute partitioning, and complete solute trapping occurs. The second IRF is given below.

\begin{align}
\label{eqn:wangIRF2}
    &v=-M\Delta G\\
    &M=\frac{v_o}{RT}\\
    &\Delta G=\sum_{i=1}^n\left[C^i_L\Delta\mu^i-\frac{V_m}{2}\alpha^i_{L}(J^i_L)^2\right]
%%    &\Delta G=\sum_{i=1}^n\left[C^i_L\Delta\mu^i-\frac{1}{2}\frac{v^2}{v_D^2}(C^i_S-C^i_L)\Delta \mu^i\right]
\end{align}

Here, in the driving force expression, first term on right-hand side is the classical driving force with solute drag effect and the second term is due to the local non-equilibrium diffusion in liquid. In the above equation, the diffusion flux of component $i$ is given by $J^i_L=(v/V_m)(C^i_L-C^i_S)$ and the non-equilibrium kinetic coefficient ($\alpha^i_L$) is given as 

\begin{equation}
    \alpha^i_L=\frac{V_m}{(v^i_D)^2}\left(\frac{\mu^i_L-\mu^i_S}{C^i_L-C^i_S}\right)
\end{equation}

In this model, the velocity-driving force relation is assumed to be linear as opposed to the exponential relation in the Ludwig model. This driving force relation accounts for the non-Fickian  solute diffusion at high solidification velocities. Now, the dendritic tip radius can be given by stability criteria as below.

\begin{align}
\label{Eqn:omega}
    &\omega=\frac{2\pi}{R}\\
    \label{Eqn:Sn}
    &S_n=-\Gamma\omega^2-(K_SG_S\xi_S+K_LG_L\xi_L)-\frac{\left(\sum_{i=2}^n\frac{M^i_L\zeta^i}{N^i}\right)}{1+\sum_{j=2}^{n}\frac{M^i_L}{N^i}}
\end{align}

According to Marginal stability criteria, the dendrite tip radius is assumed equal to the marginally stable wavelength. In Eqn.\ref{Eqn:Sn}, $G_S$ and $G_L$ are thermal gradients in solid and liquid, respectively and $K_S$ and $K_L$ are solid and liquid thermal conductivities, respectively. Other thermal variables are given below.

\begin{align}
    &\xi_S=\frac{\omega_S+v/D^T_S}{K_S\omega_S+K_L\omega_L}\\
    &\xi_L=\frac{\omega_L-v/D^T_L}{K_S\omega_S+K_L\omega_L}\\
    &\omega_L=\frac{v}{2D^T_L}+\sqrt{\left(\frac{v}{2D^T_L}\right)^2+\omega^2}\\
    &\omega_S=-\frac{v}{2D^T_S}+\sqrt{\left(\frac{v}{2D^T_S}\right)^2+\omega^2}\\
\end{align}

where $D^T_L$ and $D^T_S$ are solid and liquid thermal diffusivities respectively. Other solutal parameters are given below.
\begin{align}
\label{Eqn:kinetic_slope}
    &M^i_L=-\frac{\partial \Delta G}{\partial C^i_L}/\left[\frac{\partial \Delta G}{\partial T}+R\frac{v}{v_o}\right]\\
    &N^i=\left[D^i\psi^i\omega^i_C+vC^i_L\frac{\partial k_v^i}{\partial C^i_L}+v(k_v^i-1)\right]/vC^i_L\frac{\partial k_v^i}{\partial T}\\
    &\zeta^i=D^i\psi^iG^i_L\left[\frac{v}{D^i\psi^i}-\omega_C^i\right]/vC^i_L\frac{\partial k_v^i}{\partial T}\\
    \label{Eqn:solute_gradient}
    &G^i_L=\frac{vC^i_o}{D^i\psi^i}(1-1/k^i_v)\\
    &\omega^i_C=\frac{v}{2D^i\psi^i}+\sqrt{\left(\frac{v}{2D^i\psi^i}\right)^2+\frac{\omega^2}{\psi^i}}
\end{align}

Eqn.\ref{Eqn:kinetic_slope} represents the kinetic liquidus slope and Eqn. \ref{Eqn:solute_gradient} represents the solute gradient at the interface. In Eqn.\ref{Eqn:solute_gradient}, $C^i_o$ is the overall alloy composition. In this work, only the diagonal terms in the solute diffusion matrix are considered, and the other terms are assumed to be zero. This approximation enables faster computation and simplifies the equations \cite{wang2014morphological}. The stability equations corresponding to the diffusion equation with non-zero solute diffusion matrix can be referred to in the article by Wang et al \cite{wang2013dendritic}. 

\begin{figure}[h]
\includegraphics[width=12cm]{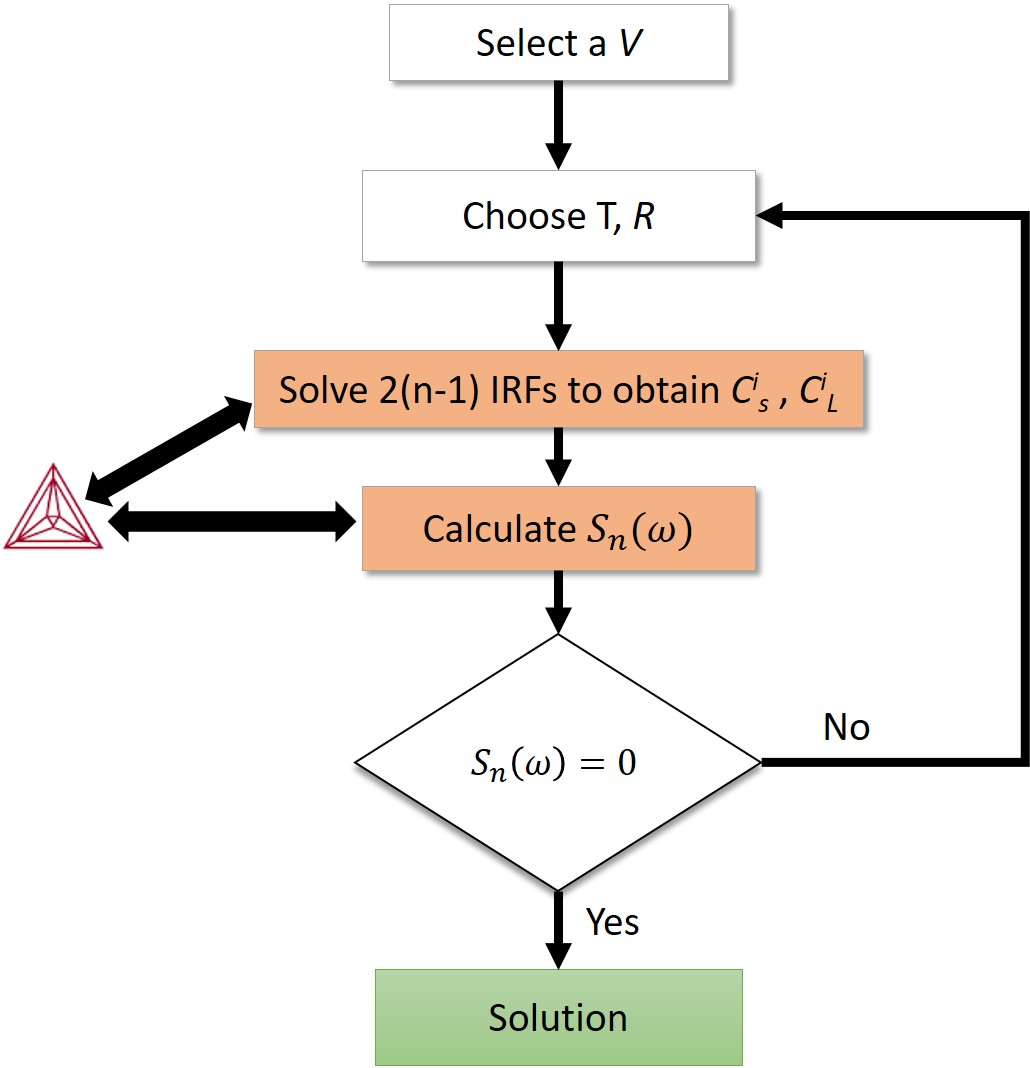}
\centering
\caption{A flowchart to calculate interface temperature and composition using Wang model. The orange colour indicates that it is Calphad coupled.}
\label{fig:Wang_schematic}
\end{figure}

In this calculation, the solid thermal gradients are assumed to be negligible compared to thermal gradients in liquid. Both solid and liquid thermal diffusivities are assumed to be the same. The diffusive speeds of all the solutes in liquid are taken as equal to that of the solvent element. The interface diffusive speeds of all elements are considered to be the same. When the thermophysical properties of solid and liquid are assumed to be same, the equations tends to be multicomponent extension of the model proposed by Li et al. \cite{li2012analysis} and in the case of dilute binary alloys, the model reduces to that of Galenko and Danilov \cite{galenko2004linear}.  A schematic representing the implementation of Wang model is shown in Fig.\ref{fig:Wang_schematic}. The thermodynamic data is obtained via TC-Python. Choose an initial velocity, and the initial guess values for radius, interface temperature and compositions can be obtained by performing a calculation using KGT model. The equations corresponding to the solid and the liquid compositions can be solved using the '\textit{fsolve}' function in '\textit{scipy}'. While calculating the interface temperature using Eqn.\ref{eqn:wangIRF2}, the temperature calculated corresponds to the planar interface and thus curvature contribution ($2\Gamma/R$) is added to account for the curved nature.  The stability and temperature equations are solved iteratively until the solution is reached. We again use '\textit{fsolve}' function of '\textit{scipy}' to arrive at the solution. Calculating kinetic liquidus slope (Eqn.\ref{Eqn:kinetic_slope}) values require computing the derivative of driving force with respect to liquid composition. Since liquid composition is a dependent variable, the driving force derivatives can not be calculated directly. We follow the below numerical procedure for the same. The velocity value is subject to a variation of  $10^{-4}$ m/s above and below the current velocity value. Using the new velocities, the compositions are obtained by solving the first IRF (Eqn:\ref{eqn:WangIRF1a}) and Ivantsov solutions (Eqn:\ref{eqn:WangIRF1b}) . Now for the new compositions, driving forces are calculated. The derivatives are now calculated using the new compositions and their corresponding driving forces.

\subsection{Columnar to equiaxed transition}
Gaumann et al. have proposed a model for columnar to equiaxed transition based on Hunt's model \cite{gaumann2001single}. This model implies that once the local undercooling reaches a critical value, all equiaxed grains nucleate in front of the columnar front, and if they are able to develop over a threshold volume fraction, they obstruct the growing columnar front. Thus, the following expression gives the thermal gradient at which a certain volume fraction of equiaxed grains nucleate.

\begin{equation}
    G=\frac{1}{n'+1}\sqrt[3]{\frac{-4\pi N_o}{3\ln(1-\phi)}}\Delta T\left(1-\frac{\Delta T_n^{n'+1}}{\Delta T^{n'+1}}\right)
\end{equation}

where $n'$ is a material-dependent parameter, $N_o$ is the nucleation volume density, $\phi$ is the volume fraction of grains,$\Delta T$ is the local undercooling and $\Delta T_n$ is the critical undercooling needed for the nucleation of equiaxed grains. Gaumann et al. neglected the kinetic, curvature and thermal contributions of undercooling and accounted only for constitutional undercooling. The constitutional undercooling was used as a fitted function of solidification velocities obtained from experiments. However, the kinetic and curvature effects are non-negligible under additive manufacturing conditions. Thus, in this work, we couple KGT model with Gaumann's model and apply for a Ni-based superalloy for LPBF conditions. Based on experimental studies, a fully equiaxed microstructure is obtained when $\phi>0.49$ and a fully columnar microstructures are obtained when $\phi<0.0066$.

The models described above require alloy properties like $T_L$, $m_v^i$, $k_v^i$, which were calculated using Thermo-Calc\textsuperscript{\textregistered}  2021B software along with TCFE11 database and other parameters such as $\Gamma, \mu_k, v_o, D_i$ were obtained from literature. For comparison, the “Scheil with solute trapping” model, which is based on Aziz and Kaplan’s model, implemented in Thermo-Calc\textsuperscript{\textregistered} is also used in the present study. 

\section{Results and Discussion}
\subsection{Case 1: Phase Selection in Fe-C-Mn-Al steel}
Babu et al., have used in situ time-resolved X-ray diffraction technique to investigate the primary solidification phase during slow and rapid cooling conditions experienced during different arc welding parameters \cite{babu2002time}. It has been observed that non-equilibrium FCC ($\gamma$) is the primary solidification phase during rapid cooling whereas; equilibrium BCC ($\delta$) is the primary solidification phase during slow cooling.Since this phase selection is observed under arc welding conditions, these calculations will be helpful for application in wire arc additive manufacturing. Thermodynamic equilibrium and IRF calculations are performed to understand and predict the experimentally observed phase selection.

\begin{figure}[h]
\includegraphics[width=16cm]{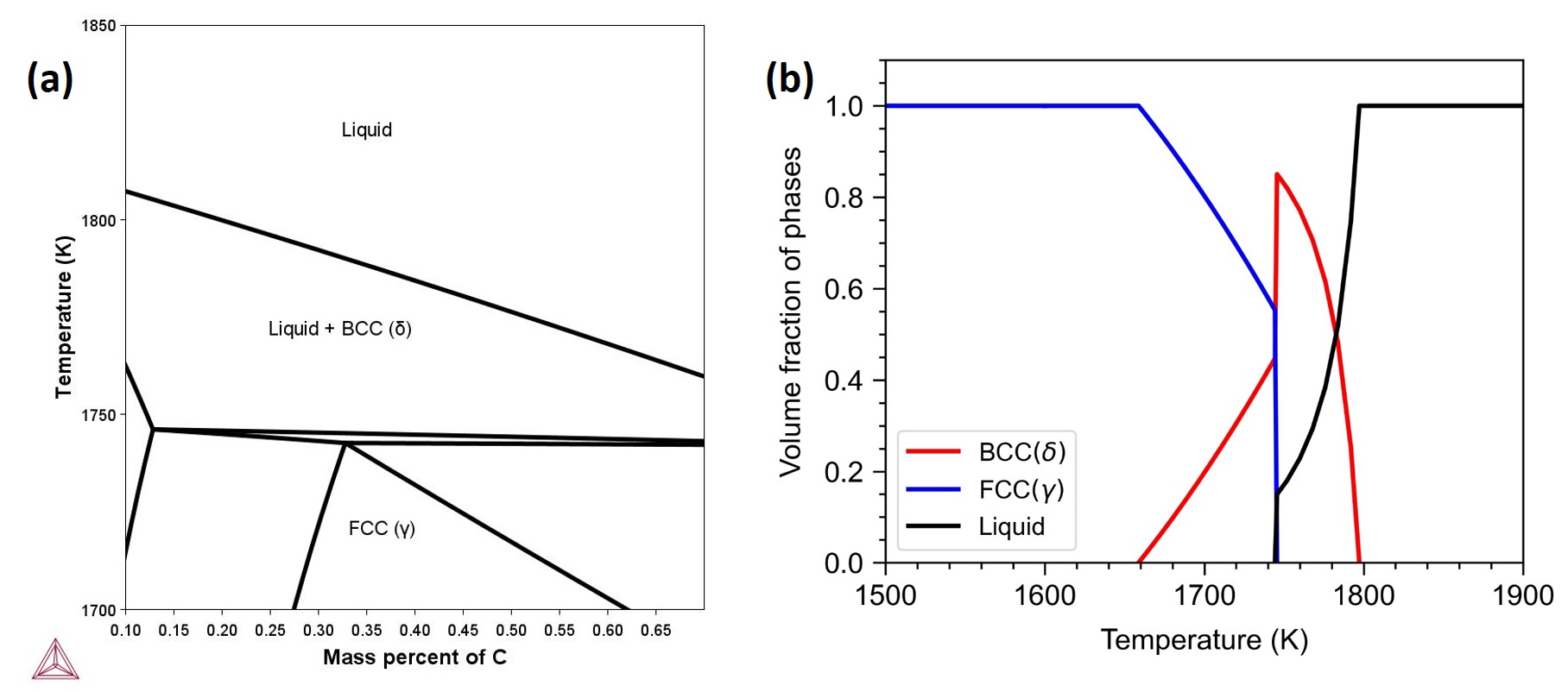}
\centering
\caption{(a) Isopleth section showing a peritectic reaction (b)Phase fraction versus temperature plot for Fe-C-Mn-Al steel}
\label{fig:Mn_steel_thermo}
\end{figure}

The composition of the Fe-C-Mn-Al steel used in the study is C-0.234, Mn-0.5, Al-1.7 and Fe-remaining (in weight percent). The isopleth section of equilibrium phase diagram is shown in Fig.\ref{fig:Mn_steel_thermo}(a), indicates the presence of the peritectic reaction (Liquid+BCC($\delta) \rightarrow \rm{FCC}(\gamma$)). The phase fraction as a function of temperature for the composition under study shown in Fig.\ref{fig:Mn_steel_thermo}(b).  From Fig.\ref{fig:Mn_steel_thermo}, it is clear that the equilibrium primary solidification phase is BCC ($\delta$). The parameters used for calculation of IRFs is given in Table\ref{tab:Mn_steel_parameters }. The value of interface diffusive speed is assumed to be slightly less than the bulk diffusive speed \cite{wang2013dendritic} The thermal gradient was considered 9$\times 10^4$ K/m \cite{wang2013dendritic}.

\begin{table}
\centering
\caption{Parameters used to calculate IRFs for Fe-C-Mn-Al steel}
\label{tab:Mn_steel_parameters }
\begin{tabular}{ccc}
\hline
\textbf{Parameter} &\textbf{BCC($\delta$)}&\textbf{FCC($\gamma$)} \\\hline
Liquidus temperature $T_L$ (K) &1796.60& 1780.98 \\
Equilibrium partition coefficient of C $k^C_o$ &0.2059&0.2716\\
Equilibrium partition coefficient of Mn $k^{Mn}_o$ &0.7129&0.7065\\
Equilibrium partition coefficient of Al $k^{Al}_o$ &1.098&0.9295\\
Equilibrium liquidus slope of C $m^C_o$ (K/wt.fraction) &-7015.3&-6569.4\\
Equilibrium liquidus slope of Mn $m^{Mn}_o$(K/wt.fraction) &-467.9&-496.7\\
Equilibrium liquidus slope of Al $m^{Al}_o$ (K/wt.fraction) &195.2&-588.7\\
Kinetic coefficient $\mu_k$ ($\rm{m s^{-1}K^{-1}}$)&10.0&10.0\\
Solute diffusion coefficient $D^i$ ($\rm{m^2/s}$)&$5.0\times10^{-9}$&$5.0\times10^{-9}$\\
Diffusive speed in liquid $v_D$ (m/s)& 1.0& 1.0\\
max speed $v_o$ (m/s)&300.0&300.0\\
Interface diffusion speed $v_I$ (m/s)&0.9&0.9\\
Gibbs-Thomson coefficient $\Gamma$ (K/m)&$2.0\times10^{-7}$&$1.0\times10^{-7}$\\
Solid and Liquid Thermal conductivity $K_S,K_L$ ($\rm{Wm^{-1}K^{-1}}$) &50&50 \\
Solid and Liquid Thermal Diffusivity$D^T_S,D^T_L$  ($\rm{m^2s^{-1}}$) &$5.0\times10^{-5}$&$5.0\times10^{-5}$ \\
\hline
\end{tabular}
\end{table}

\begin{figure}[h]
\includegraphics[width=16cm]{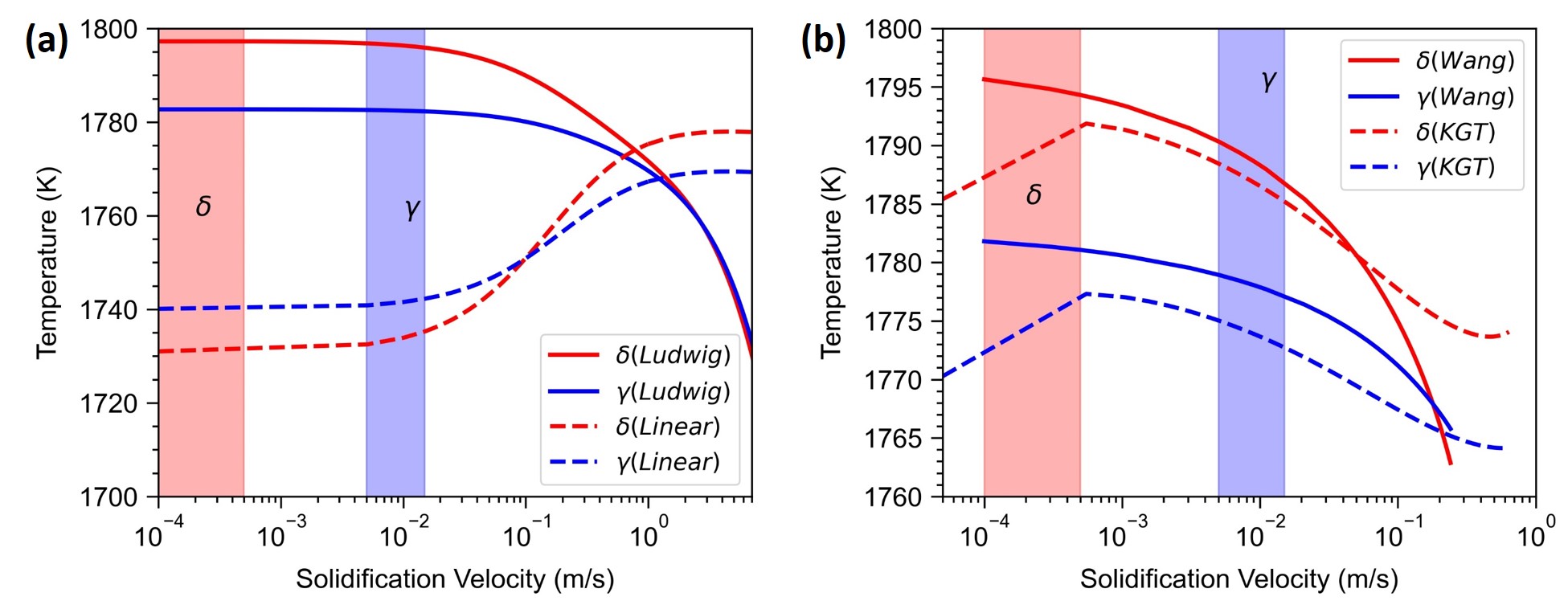}
\centering
\caption{The vertical coloured areas indicate the experimental observations. Plot of interface temperature for BCC($\delta$) and FCC($\gamma$) solidification for (a) planar models (b)dendritic models}
\label{fig:Mn_steel_IRF}
\end{figure}

The BCC($\delta$)-liquid and FCC($\gamma$)-liquid interface temperature as a function of solidification velocity for all the above mentioned models are shown in Fig.\ref{fig:Mn_steel_IRF}. The planar solidification models with linear phase diagram shows increase in the interface temperature with increasing solidification velocity, for both BCC and FCC. However, the interface temperature is observed to decrease for all other models. In the case of Ludwig's model, BCC phase has higher interface temperature than FCC for low velocities, whereas at high velocities, FCC has higher interface temperature. The phase with higher interface temperature will be the primary solidification phase \cite{kurz2001solidification}. Thus, Ludwig's model qualitatively predicts the phase selection that is experimentally observed during welding. However, it overestimates the velocity at which the transition occurs. Ludwig's model shown in Fig.\ref{fig:Mn_steel_IRF}(a) takes solute drag into account. In the case of dendritic models, KGT model does not show any phase selection whereas, Wang's model shows FCC as the primary solidification phase for velocities greater than ~1 cm/s. Compared to Ludwig's model, Wang's model predicts a lower velocity at which the transition occurs.

In both planar and curved interface models, the coupling of thermodynamic data has improved the predictions. KGT model will not predict the phase transition even if the partition coefficients and liquidus slopes are directly calculated from calphad models for every iteration \cite{babu2002time}. Mohan and Phanikumar showed that the sum of the constitutional undercooling contribution of individual solutes is less than the constitutional undercooling values calculated via Calphad databases using the liquid composition \cite{mohan2019experimental}. The explicit addition of individual undercoolings is valid only for dilute binary alloys but for non-dilute alloys, the method is shown to be invalid by Ludwig \cite{ludwig1998interface}. Since both Ludwig's and Wang's models account for the thermodynamic interactions among the elements, the phase transition is being captured. As Wang's model considers additional physics, such as the curved nature of the interface and non-Fickian diffusion, it can predict closer to experiments than the case of Ludwig's model.

\begin{figure}
    \centering
    \includegraphics[width=16cm]{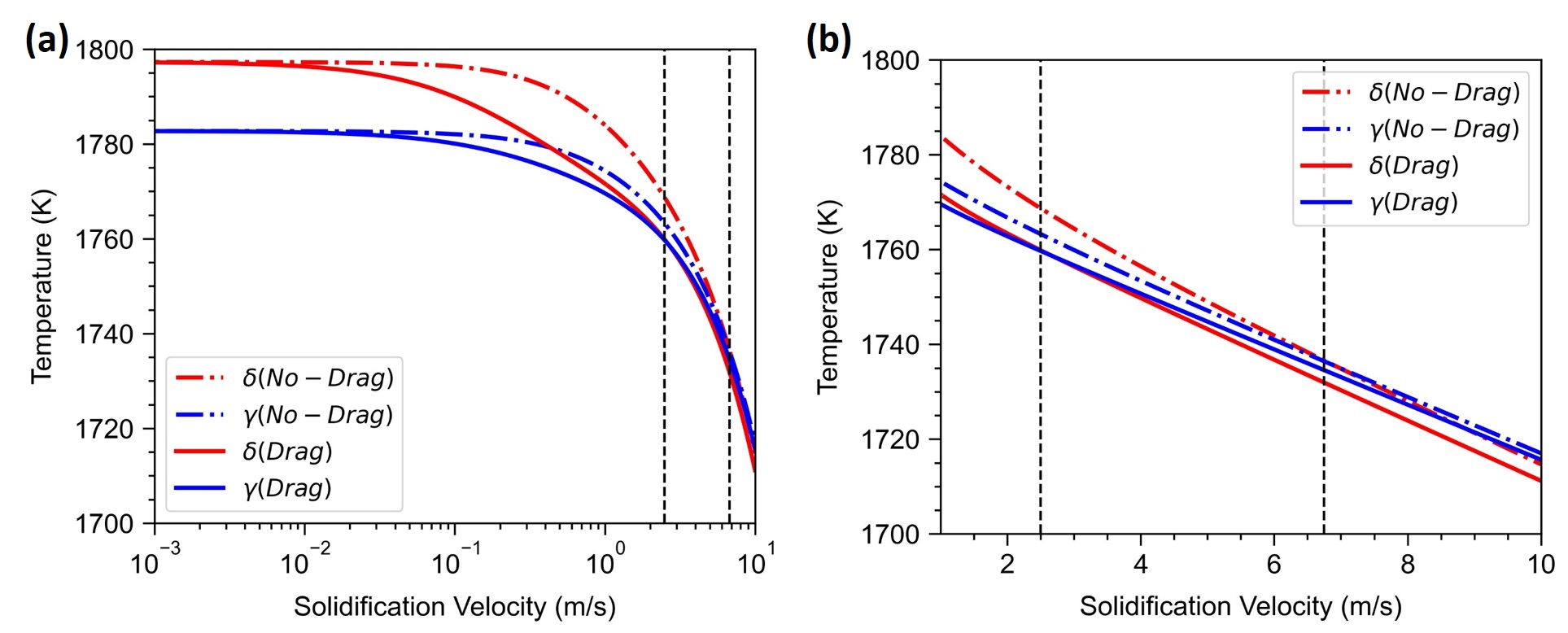}
    \caption{(a)Interface temperature versus solidification velocity calculated using Ludwig's model with and without solute trapping for Fe-C-Mn-Al steel (b)Magnified view showing the transition}
    \label{fig:Mn_IRF2}
\end{figure}

\begin{figure}[h]
    \centering
    \includegraphics[width=12cm]{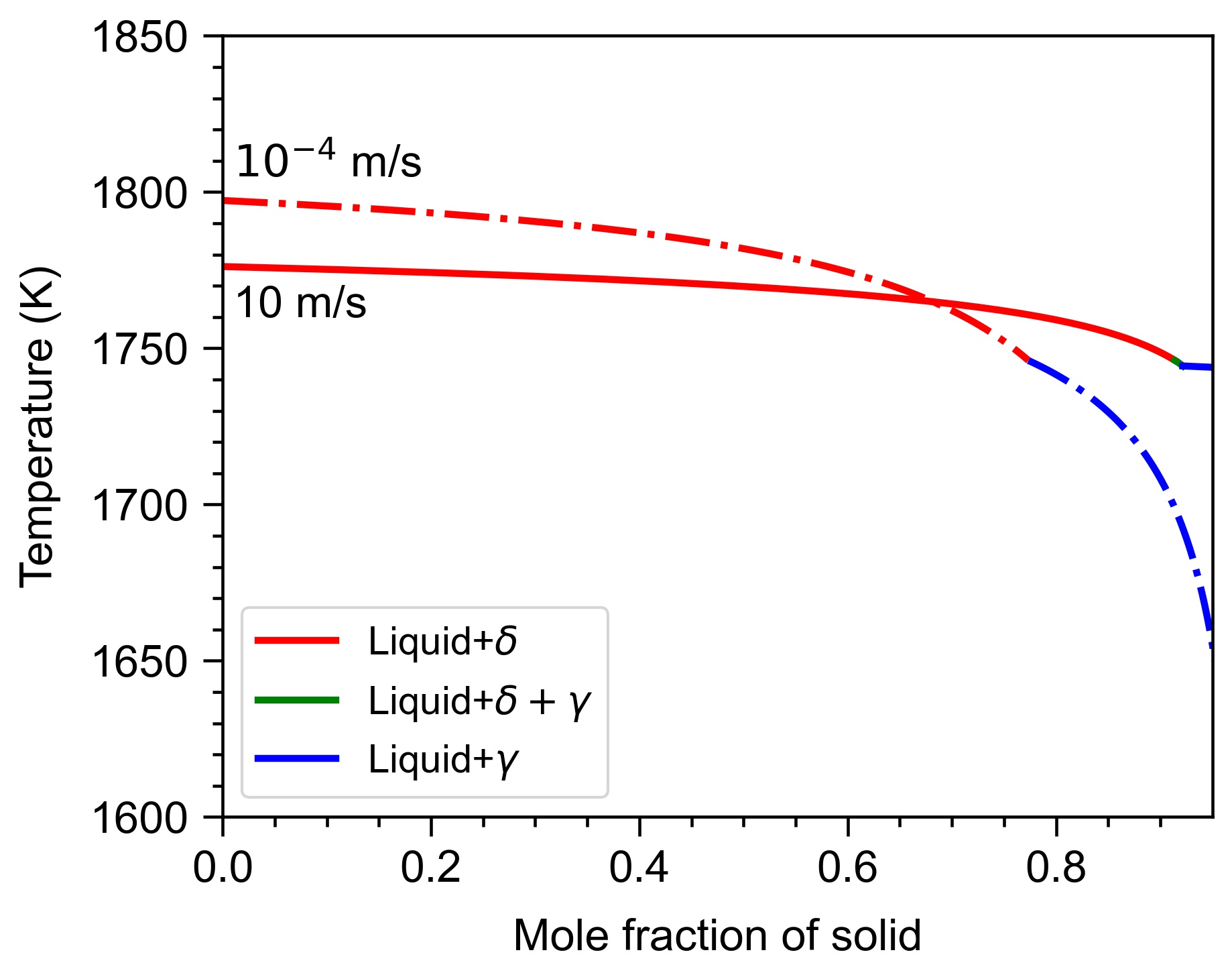}
    \caption{The temperature versus fraction of solid calculated using Thermo-Calc 'Scheil with solute trapping model' for two extreme solidification velocities}
    \label{fig:Mn_steel_thermocalc}
\end{figure}

The effect of solute drag on the interface temperature is shown in Fig.\ref{fig:Mn_IRF2}. Including the solute drag has improved the prediction and the transition occurs at a lower velocity. As a part of the driving force is consumed due to solute transfer across the interface, there is a decrease in driving force available for interface motion, leading to decreased velocity. The temperature versus fraction of solid plot for Fe-C-Mn-Al steel calculated using Thermo-Calc\textsuperscript{\textregistered}'s 'Scheil with solute trapping model' is shown in Fig.\ref{fig:Mn_steel_thermocalc}. Since it is planar, we will compare the results of Ludwig's model with Thermo-Calc's value.Note the temperature when fraction of solid is zero. When the solidification velocity is increased, there is a decrease in the interface temperature as expected. This observation matches qualitatively with our calculation using Ludwig's model, although there is a quantitative difference. However, the primary solidification phase is BCC even at such high solidification velocities. This is in contrast with the experimental data. Thus, Ludwig's model and Wang's model are able to show the experimentally observed phase selection but not other IRF models.

\subsection{Case 2: Phase Selection in H13 tool steel}
Konig et al. have carried out in situ synchrotron-based high speed X ray diffraction experiments in different LPBF conditions to understand the primary solidification modes of H13 tool steels during LPBF \cite{konig2023solidification}. BCC-$\delta$ ferrite was the primary solidification phase when cooled with a cooling rate of $\sim 2\times 10^4$ K/s and FCC-$\gamma$ austenite was the primary solidification phase when cooled with a higher cooling rate of $\sim 1\times10^6$ K/s. Thermodynamic and IRF calculations are performed to explain this phase selection.

\begin{figure}[h]
\includegraphics[width=16cm]{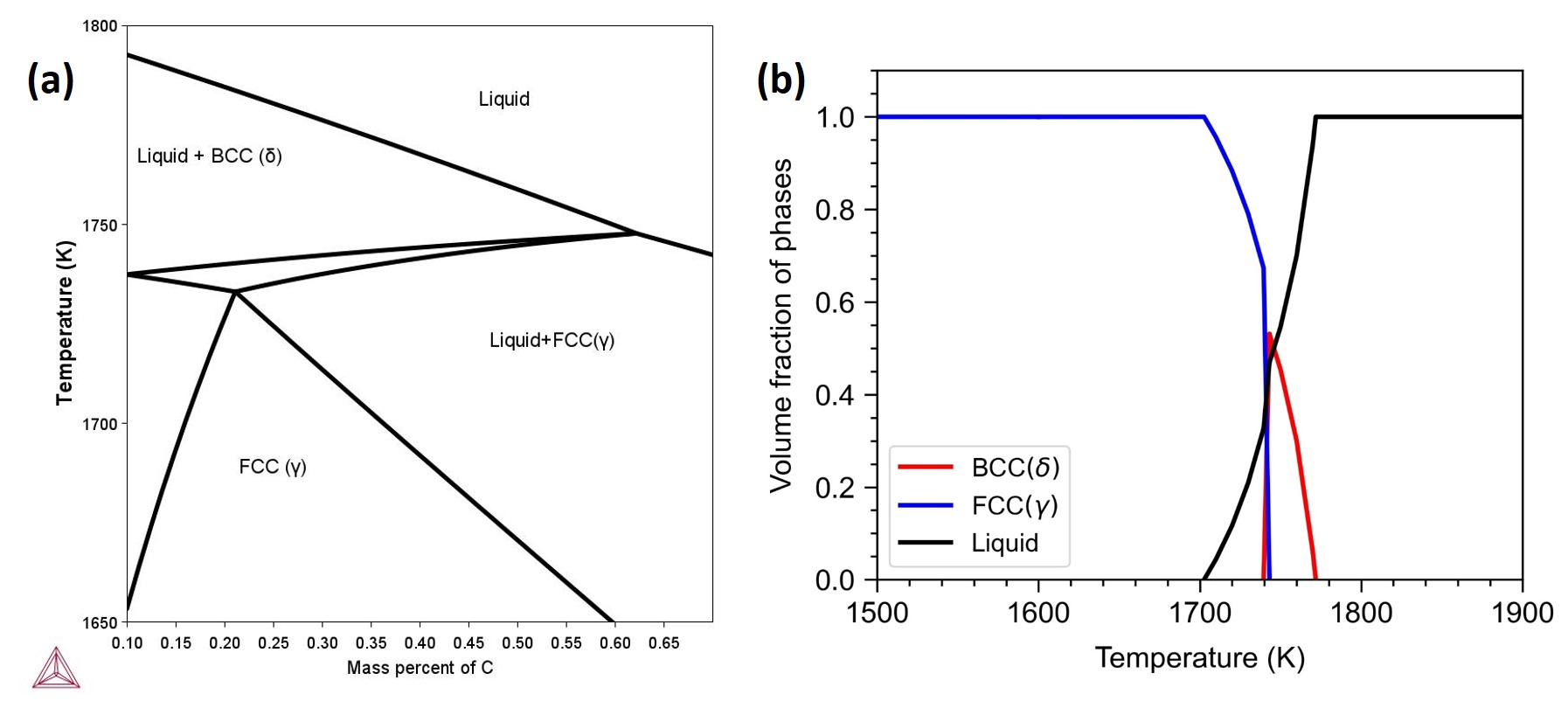}
\centering
\caption{(a) Isopleth section showing a peritectic reaction (b)Phase fraction versus temperature plot for H13 tool steel}
\label{fig:H13_steel_thermo}
\end{figure}

The composition used in the study is C-0.35, Mo-2.24, Mn-0.45, Si-0.25, Fe-Remaining (weight percent). The isopleth diagram shown in Fig.\ref{fig:H13_steel_thermo}(a) reveals the presence of peritectic reaction. The phase fraction versus temperature plot in Fig.\ref{fig:H13_steel_thermo}(b) corroborates that for the considered composition, BCC($\delta$) is the primary solidification phase under equilibrium conditions. The parameters used for IRF calculations are given in Table \ref{tab:H13_steel_parameters}. The thermal gradient is taken as 1$\times 10^5$ K/m, which is typical for LPBF. 

\begin{table}
\centering
\caption{Parameters used to calculate IRFs for H13 tool steel}
\label{tab:H13_steel_parameters}
\begin{tabular}{ccc}
\hline
\textbf{Parameter} &\textbf{BCC($\delta$)}&\textbf{FCC($\gamma$)} \\\hline
Liquidus temperature $T_L$ (K) &1771.89& 1765.78 \\
Equilibrium partition coefficient of C $k^C_o$ &0.156&0.2943\\
Equilibrium partition coefficient of Mo $k^{Mo}_o$ &0.723&0.6118\\
Equilibrium partition coefficient of Mn $k^{Mn}_o$ &0.718&0.7564\\
Equilibrium partition coefficient of Si $k^{Si}_o$ &0.672&0.6430\\
Equilibrium liquidus slope of C $m^C_o$ (K/wt.fraction) &-8580.037&-6470.758\\
Equilibrium liquidus slope of Mo $m^{Mo}_o$(K/wt.fraction) &-261.256&-382.766\\
Equilibrium liquidus slope of Mn $m^{Mn}_o$ (K/wt.fraction) &-521.308&-411.965\\
Equilibrium liquidus slope of Si $m^{Si}_o$ (K/wt.fraction) &-1256.316&-1315.602\\
Kinetic coefficient $\mu_k$ ($\rm{m s^{-1}K^{-1}}$)&10.0&10.0\\
Solute diffusion coefficient $D^i$ ($\rm{m^2/s}$)&$5.0\times10^{-9}$&$5.0\times10^{-9}$\\
Diffusive speed in liquid $v_D$ (m/s)& 1.0& 1.0\\
max speed $v_o$ (m/s)&300.0&300.0\\
Interface diffusion speed $v_I$ (m/s)&0.9&0.9\\
Gibbs-Thomson coefficient $\Gamma$ (K/m)&$1.0\times10^{-7}$&$5.0\times10^{-8}$\\
Solid and Liquid Thermal conductivity $K_S,K_L$ ($\rm{Wm^{-1}K^{-1}}$) &50&50 \\
Solid and Liquid Thermal Diffusivity$D^T_S,D^T_L$  ($\rm{m^2s^{-1}}$) &$5.0\times10^{-5}$&$5.0\times10^{-5}$ \\
\hline
\end{tabular}
\end{table}

\begin{figure}[h]
\includegraphics[width=16cm]{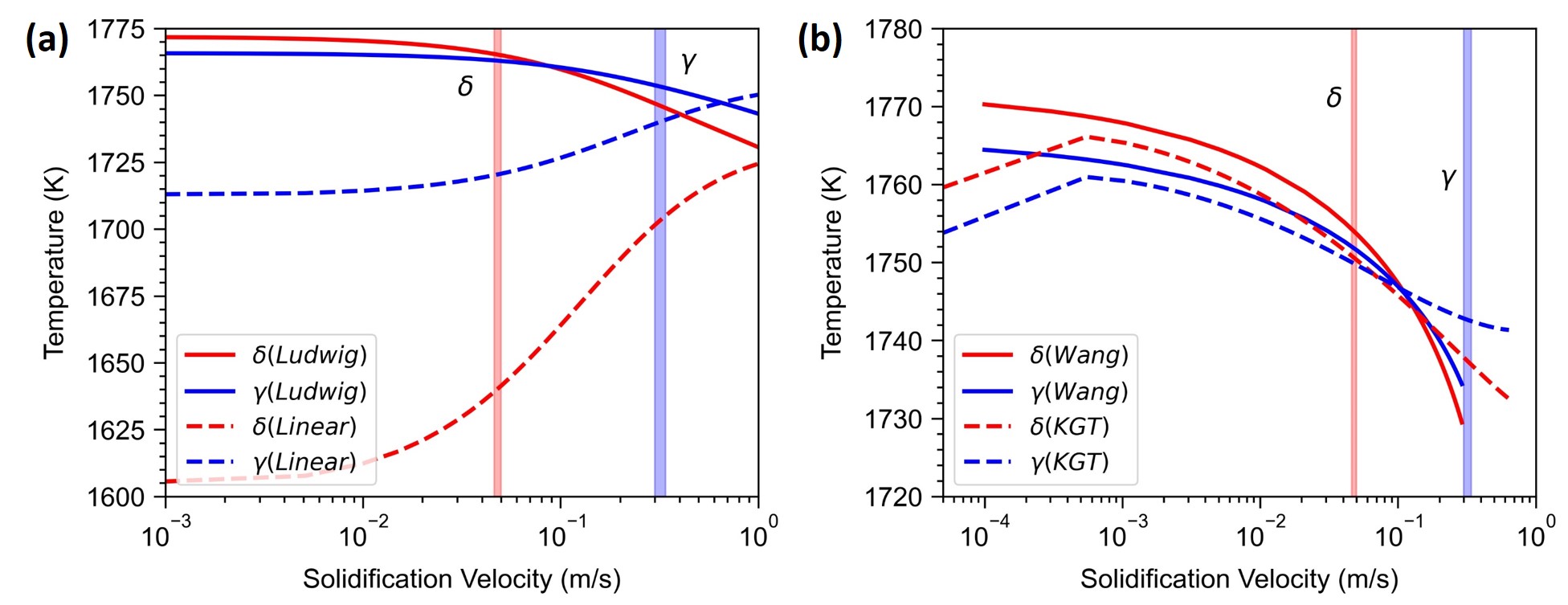}
\centering
\caption{The vertical coloured areas indicate the experimental observations. Plot of interface temperature for BCC($\delta$) and FCC($\gamma$) solidification for (a) planar models (b)dendritic models}
\label{fig:H13_steel_IRF}
\end{figure}

The variation in the interface temperature with solidification velocity is shown in Fig.\ref{fig:H13_steel_IRF}. For both BCC and FCC, the planar solidification models with linear phase diagram predict a rise in interface temperature with increasing solidification velocity. But, the interface temperature decreases for all other models. This is similar to that observed in Case 1. Except planar model with linear phase diagram, all other models predicted the phase selection and the transition velocities predicted are very close to the experimentally observed values.

\begin{figure}[h]
    \centering
    \includegraphics[width=16cm]{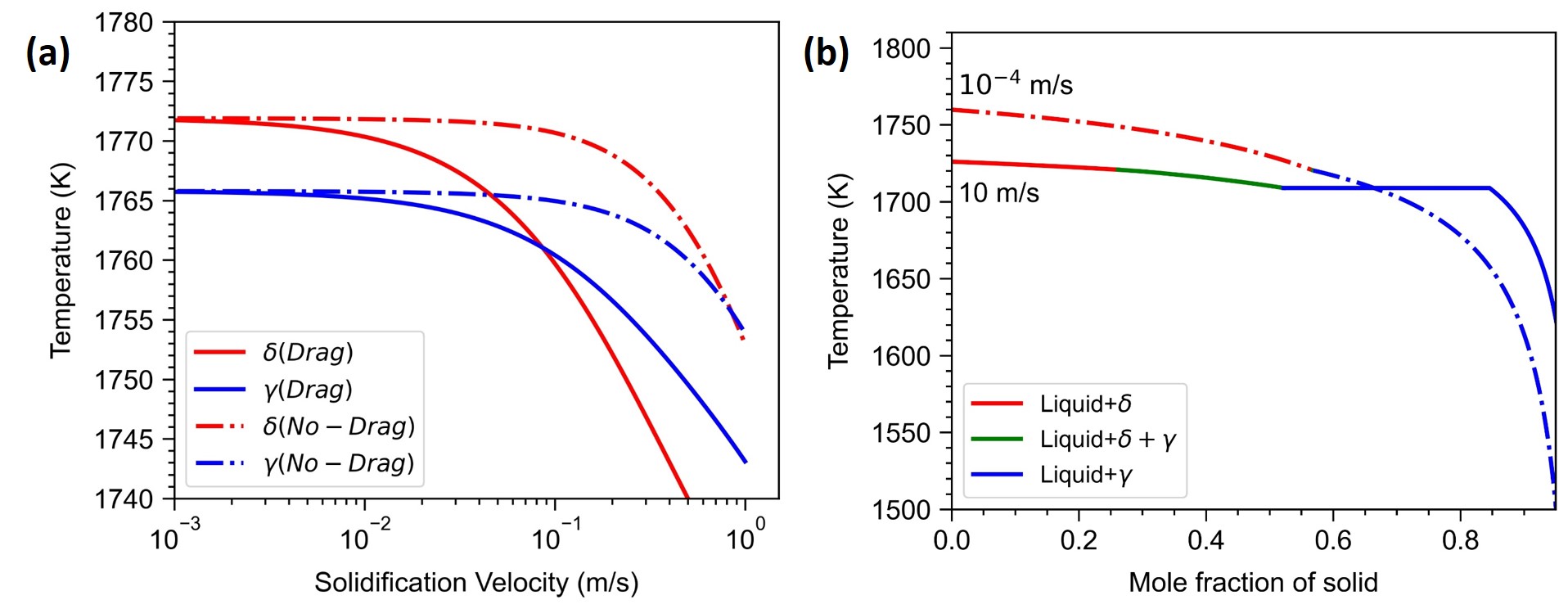}
    \caption{(a)Interface temperature versus solidification velocity calculated using Ludwig's model with and without solute trapping for H13 tool steel (b)The temperature versus fraction of solid calculated using Thermo-Calc 'Scheil with solute trapping model' for two extreme solidification velocities}
    \label{fig:H13_IRF2}
\end{figure}

Fig.\ref{fig:H13_IRF2}(a) depicts the influence of solute drag on interface temperature. Adding solute drag improved the prediction, and the changeover occurred at a lower velocity. The effect of solute drag is more drastic in this case than in the previous one. A plausible explanation for such a pronounced effect might be the increased concentration of solute elements like Mo and Si that partition to liquid. When there is severe partitioning of elements into liquid at the interface, the compositional difference between the solid and the liquid at the interface increases leading to an increase in the Gibbs energy dissipation due to the solute drag (in Eqn.\ref{Eqn:Solute_Drag_Energy}). This leads to decrease in solidification velocity. A similar effect has been observed in  FeCoNiCuSn high entropy alloy when concentration of Sn was increased \cite{rahul2020growth}. In Case 1, Al has a partition coefficient close to 1.0, thus the interfacial composition difference is minimal and the solute drag effect is not very predominant. The temperature versus fraction of solid plot for H13 steel calculated using Thermo-Calc\textsuperscript{\textregistered}'s 'Scheil with solute trapping model' is shown in Fig.\ref{fig:H13_IRF2}(b). The initial temperature (when fraction of solid is zero) decreases with increase in solidification velocity. This observation agrees with the calculations from Ludwig's model. However, it predicts $\delta$ as the primary solidification phase in both extreme velocities, contradicting the experimental observation.\par

From the above cases, it is clear that the thermodynamic coupling of IRFs improved the predictions. Additionally, including the effect of solute drag increased the accuracy. Instead of either including or excluding solute drag, one can use a partial drag parameter to match with the experiments and such a model is proposed by Hareland et al. \cite{hareland2022thermodynamics}.  In this work, non-diagonal terms in diffusion matrix are neglected. Guillemot et al. have highlighted the importance of non-diagonal terms in diffusion matrix and used Hunziker's algorithm to account for the cross-diffusion of solutes \cite{guillemot2022thermodynamic,hunziker2001theory}. From their report, it is clear that even in a concentrated eight component solid solution, neglecting the effect of cross-diffusion leads to a difference of 10 K for velocities close to 1 m/s (Fig. 7(b) in Ref. \cite{guillemot2022thermodynamic}). Thus, neglecting the cross-diffusion effects seems to be a reasonable approximation. Sobolev et al. recently highlighted the effect of cross-diffusion terms when hyperbolic type (non-Fickian) of diffusion equation is considered for the re-solidification occurring during femtosecond laser pulse melting where the solidification velocities are much higher \cite{sobolev2023rapid}. The deviations are expected to be much less for the range of velocities considered in our study. Another approximation used in both the dendritic models is that the marginally stable wavelength is selected as the dendritic tip radius (Marginal Stability criterion). Such an assumption was made based on experimental data and the physics behind such selection is lagging in marginal stability criterion. Microsolvability criterion fills that gap by taking anisotropic effects into account and has been verified by phase-field simulations \cite{kurz2019progress,karma1997phase}. In Marginal stability criterion, the selection parameter takes the value of $(2\pi)^{-2}$, but in Microsolvability theory, this parameter depends on the anisotropy. One can calibrate the selection parameter using selected phase-field simulations \cite{lahiri2019dendrite}, but it is beyond the scope of the current work.\par

In spite of these limitations, the above-mentioned models can serve as important tools to understand the kinetically controlled phase selections occurring in metal additive manufacturing.

\subsection{Case 3: Columnar to Equiaxed Transition in Haynes 282}

\begin{figure}[h]
    \centering
    \includegraphics[width=12cm]{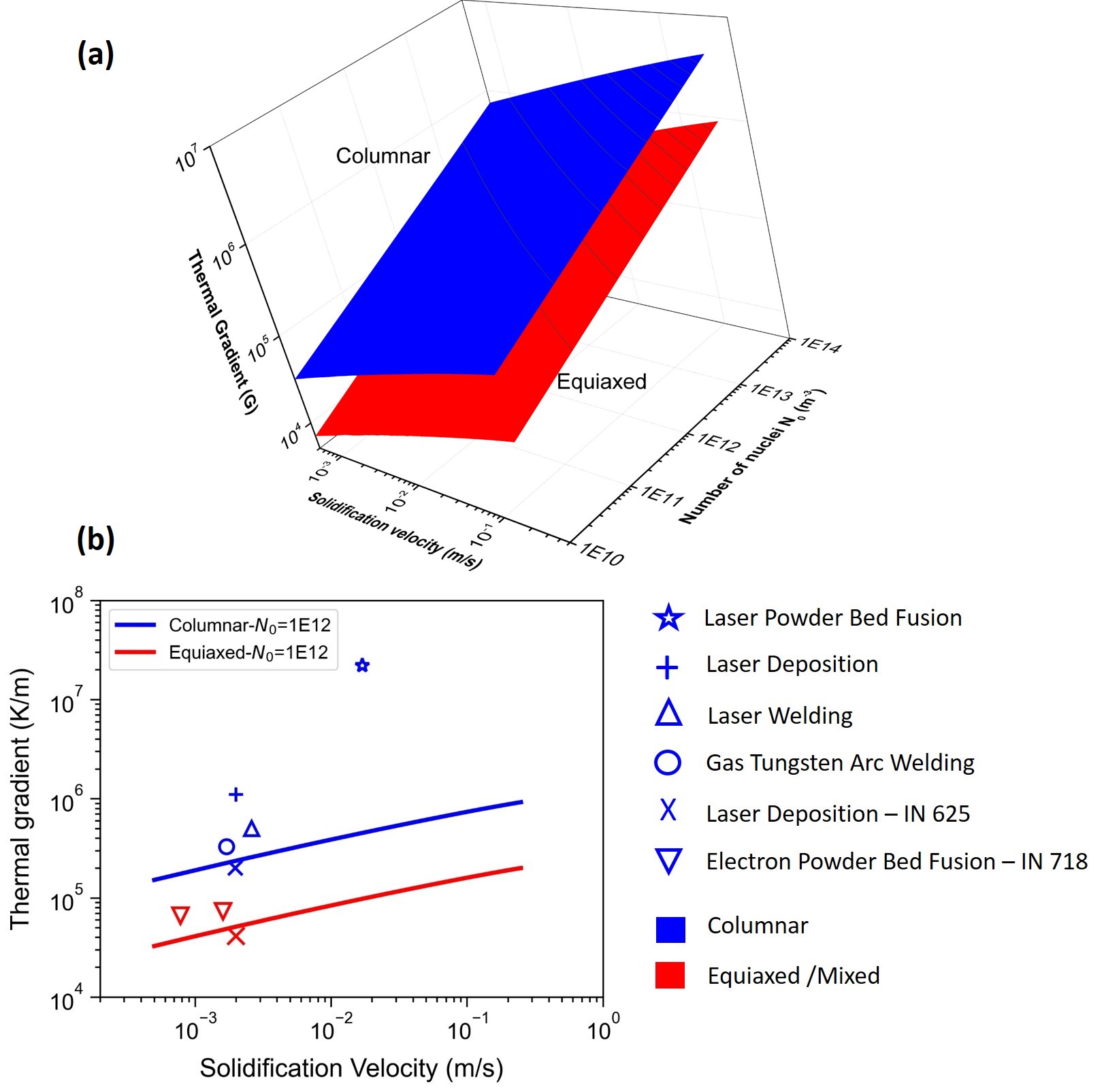}
    \caption{(a)Variation in CET plot for Haynes 282 with varying $N_o$ (b)CET plot corresponding to $N_o=1\times10^{12}$ along with experimental data from LPBF \cite{hariharan2022icme}, Laser deposition \cite{ramakrishnan2019microstructure},Laser welding \cite{osoba2012microstructural},Gas tungsten arc welding \cite{sivaji2023},Laser deposition of IN625 \cite{li2022controlling}, Electron Powder Bed Fusion \cite{raghavan2021influence}}
    \label{fig:Haynes282_CET}
\end{figure}

Haynes 282 is a $\gamma'$ strengthened Ni-based superalloy designed for high temperature strength and thermal stability without compromising fabricability. Additive manufacturing of Haynes 282 is gaining attention as the additively manufactured components have better mechanical properties than the wrought alloys. Haynes 282 has been successfully fabricated using electron powder bed fusion (EPBF) \cite{unocic2020evaluation}, LPBF \cite{hariharan2022icme}, laser deposition \cite{ramakrishnan2019microstructure} and wire arc additive manufacturing \cite{zhang2022evolution}. Fernandez-Zelaia et al. have successfully fabricated single crystalline Haynes 282 using EPBF \cite{fernandez2021nickel}. Thus, understanding and predicting CET becomes necessary. Here, CET plot is obtained by iteratively solving KGT model for each solidification velocity. The parameters used to calculate KGT model is given in Table \ref{tab:Haynes282_KGT}. The critical undercooling for nucleation ($\Delta T_n$) is taken as zero due to the epitaxial nature of solidification commonly observed during welding and AM \cite{gaumann2001single}.

\begin{table}
\centering
\caption{Parameters used to calculate IRF (KGT model) for Haynes 282}
\label{tab:Haynes282_KGT}
\begin{tabular}{cc}
\hline
\textbf{Parameter} &\textbf{Value} \\\hline
Liquidus temperature $T_L$ (K) &1641.99\\
Equilibrium partition coefficient of Cr $k^{Cr}_o$ &1.005\\
Equilibrium partition coefficient of Co $k^{Co}_o$ &1.144\\
Equilibrium partition coefficient of Mo $k^{Mo}_o$ &0.788\\
Equilibrium partition coefficient of Ti $k^{Ti}_o$ &0.429\\
Equilibrium partition coefficient of Al $k^{Al}_o$ &1.002\\
Equilibrium liquidus slope of Cr $m^{Cr}_o$ (K/wt.fraction) &-315.687\\
Equilibrium liquidus slope of Co $m^{Co}_o$(K/wt.fraction) &82.478\\
Equilibrium liquidus slope of Mo $m^{Mo}_o$ (K/wt.fraction) &-432.653\\
Equilibrium liquidus slope of Ti $m^{Ti}_o$ (K/wt.fraction) &-1786.874\\
Equilibrium liquidus slope of Al $m^{Al}_o$ (K/wt.fraction) &-778.161\\
Kinetic coefficient $\mu_k$ ($\rm{m s^{-1}K^{-1}}$)&10.0\\
Solute diffusion coefficient $D^{Cr}$ ($\rm{m^2/s}$)&$4.0\times10^{-9}$\\
Solute diffusion coefficient $D^{Co}$ ($\rm{m^2/s}$)&$2.8\times10^{-9}$\\
Solute diffusion coefficient $D^{Mo}$ ($\rm{m^2/s}$)&$1.3\times10^{-9}$\\
Solute diffusion coefficient $D^{Ti}$ ($\rm{m^2/s}$)&$4.7\times10^{-9}$\\
Solute diffusion coefficient $D^{Al}$ ($\rm{m^2/s}$)&$3.8\times10^{-9}$\\
Diffusive speed in liquid $v_D$ (m/s)& 10.0\\
max speed $v_o$ (m/s)&300.0\\
Gibbs-Thomson coefficient $\Gamma$ (K/m)&$1.0\times10^{-7}$\\
Material parameter $n'$&2.0\\
\hline
\end{tabular}
\end{table}

The variation in CET plot for Haynes 282 with varying $N_o$ is shown in Fig.\ref{fig:Haynes282_CET}(a). Haines et al. studied the sensitivity of different variables on the CET plot of IN 718 alloy and have demonstrated that the $N_o$ is the most sensitive parameter that causes significant changes to the CET plot \cite{haines2018sensitivity}. Based on the experimental evidence from different processing conditions, one has to choose the $N_o$ and it has been treated as a fitting parameter in several studies \cite{gaumann2001single,haines2018sensitivity}. The experimentally observed microstructures are plotted along with the calculated CET plot for $N_o=1\times10^{12}\rm{m^{-3}}$ in Fig.\ref{fig:Haynes282_CET}(b). For the process conditions reported in the literature, the thermal gradient and solidification velocities are estimated using the Rosenthal solution \cite{promoppatum2017comprehensive}. Even though there are several approximations like ignoring convection within the liquid pool and assuming constant thermophysical properties, it gives an estimate of the processing conditions for us to perform CET or IRF calculations. Due to the limited availability of data related to Haynes 282, other related Ni-based superalloys like IN 625 and IN 718 are used for comparison.

When single crystal or columnar microstructure are preferred, one has to choose the process conditions, such that the gradients and velocities lie above the columnar (blue) line. When equiaxed microstructures are preferred, process conditions must be chosen such that they are in the region below the equiaxed (red) line. In laser powder deposition, the solidification front at the top of the melt pool experiences low thermal gradients and high solidification velocities. Hence, equiaxed grains are commonly observed at the top of the melt pool \cite{kurz2001columnar}. In EPBF, one can obtain either columnar or mixed microstructure by adjusting the process conditions or scan patterns \cite{fernandez2020crystallographic}. Whereas in LPBF, the thermal gradients are generally higher and difficult to achieve equiaxed microstructure only by process parameter modification. External particles or inoculants that increase the $N_o$ value are added to induce heterogeneous nucleation and resulting in equiaxed microstructure \cite{haines2018sensitivity,WANG2022102615}.\par

\section{Conclusions}
The study demonstrated the capability of Calphad coupled IRFs to predict the phase selection during AM, that are otherwise difficult by models based on linear superposition of phase diagram. The thermodynamic interactions among the elements are ignored in models with linear superposition, which can become non-negligible for multicomponent alloys and lead to erroneous results. The methods described here allow the users to have a quick calculation for the phase selection. These calculations can serve as preliminary results before performing more sophisticated phase-field calculations. These models provide qualitative trends and insights, thus serving as a guide for both process and alloy design for additive manufacturing.

\section*{CRediT authorship contribution statement}
\textbf{V S Hariharan:} Conceptualization, Data Curation, Software, Writing, Formal Analysis. \textbf{B S Murty:} Investigation, Reviewing and Editing, \textbf{G Phanikumar:} Conceptualization, Formal Analysis, Funding, Reviewing and Editing.

\section*{Declaration of Competing Interest}
The authors declare that they have no known competing financial interests or personal relationships that could have appeared to influence the work reported in this paper.

\section*{Acknowledgement}
Hariharan thanks Dr. Dasari Mohan (IIT Bombay, India) for his technical discussions and inputs which helped improve this work. Hariharan thanks Dr. Kang Wang (University of Virginia, USA) for his suggestions in implementing Wang model.

\section*{Data Availability}

\href{https://github.com/ICME-India/solidification-IRF}{https://github.com/ICME-India/solidification-IRF}

\bibliographystyle{elsarticle-num} 
\bibliography{cas-refs}





% \end{thebibliography}
\end{document}